\documentclass[conference]{IEEEtran}
\IEEEoverridecommandlockouts

\usepackage{amsmath}
\usepackage{amsfonts}
\usepackage{lipsum}
\usepackage{graphicx}
\usepackage{xspace}
\usepackage{epsf}
\usepackage{setspace}
\usepackage{float}
\usepackage{listings}
\usepackage{tikz}
\usepackage{extdash}
\usepackage{textcomp}
\usepackage{fancyhdr}

\usepackage[]{siunitx}
\DeclareSIUnit\flop{FLOP}
\usepackage{microtype}
\usepackage[numbers,square,sort&compress]{natbib}

\usepackage{soul}
\setstcolor{red}

\usepackage[margin=1in]{geometry}

\usepackage{stmaryrd}
\SetSymbolFont{stmry}{bold}{U}{stmry}{m}{n}

\usepackage[]{xcolor}
\usepackage{bm}
\newcommand{\ve}[1]{\bm{#1}}

\newcommand{\bu}{\ve{u}}

\newcommand{\bq}{\ve{q}}

\newcommand{\bh}{\ve{h}}

\newcommand{\bA}{\ve{A}}

\newcommand{\bF}{\ve{F}}
\newcommand{\bI}{\ve{I}}

\definecolor{lightblue}{rgb}{0.63, 0.74, 0.78}
\definecolor{seagreen}{rgb}{0.18, 0.42, 0.41}
\definecolor{orange}{rgb}{0.85, 0.55, 0.13}
\definecolor{silver}{rgb}{0.69, 0.67, 0.66}
\definecolor{rust}{rgb}{0.72, 0.26, 0.06}
\definecolor{purp}{RGB}{68, 14, 156}

\colorlet{lightrust}{rust!50!white}
\colorlet{lightorange}{orange!25!white}
\colorlet{lightlightblue}{lightblue}
\colorlet{lightsilver}{silver!30!white}
\colorlet{darkorange}{orange!75!black}
\colorlet{darksilver}{silver!65!black}
\colorlet{darklightblue}{lightblue!65!black}
\colorlet{darkrust}{rust!85!black}
\colorlet{darkseagreen}{seagreen!85!black}

\usepackage{hyperref}
\hypersetup{
  colorlinks=true,
}

\usepackage[labelfont=bf,font=small]{caption}

\usepackage{cleveref}
\crefname{lstlisting}{listing}{listings}
\Crefname{lstlisting}{Listing}{Listings}

\definecolor{dkgreen}{rgb}{0,0.6,0}
\definecolor{gray}{rgb}{0.5,0.5,0.5}
\definecolor{mauve}{rgb}{0.58,0,0.82}

\usepackage{parskip}

\usepackage{xparse}
\DeclareDocumentCommand{\diff}{O{} m}{
	\frac{\mathrm{d} #1}{\mathrm{d}#2}
}
\DeclareDocumentCommand{\difftwo}{O{} m}{
	\frac{\mathrm{d}^2 #1}{\mathrm{d}#2^2}
}
\DeclareDocumentCommand{\pdiff}{O{} m}{
	\frac{\partial #1}{\partial #2}
}
\DeclareDocumentCommand{\pdifftwo}{O{} m}{
	\frac{\partial^{2} #1}{\partial #2^{2}}
}
\DeclareDocumentCommand{\integral}{O{} O{} m O{x}}{
	\int_{#1}^{#2} #3\ \mathrm{d}#4
}

\usepackage{extdash}

\hypersetup{
  linkcolor=darkrust,
  citecolor=seagreen,
  urlcolor=darkrust,
  pdfauthor=author,
}

\def\BibTeX{{\rm B\kern-.05em{\sc i\kern-.025em b}\kern-.08em
    T\kern-.1667em\lower.7ex\hbox{E}\kern-.125emX}}
    
\begin{document}

\title{\huge{OpenACC offloading of the MFC compressible multiphase flow solver on AMD and NVIDIA GPUs}
}

\author{\IEEEauthorblockN{Benjamin Wilfong}
\IEEEauthorblockA{\textit{Computational Science and Eng.} \\
\textit{Georgia Institute of Technology}\\
Atlanta, Georgia} \\
\IEEEauthorblockN{Steve Abbott}
\IEEEauthorblockA{\textit{Hewlett Packard Enterprise} \\
Bloomington, Minnesota}
\and
\IEEEauthorblockN{Anand Radhakrishnan}
\IEEEauthorblockA{\textit{Computational Science and Eng.} \\
\textit{Georgia Institute of Technology}\\
Atlanta, Georgia} \\
\IEEEauthorblockN{Reuben D. Budiardja}
\IEEEauthorblockA{
\textit{Oak Ridge National Laboratory}\\
Oak Ridge, Tennessee \\
}
\and
\IEEEauthorblockN{Henry A. Le Berre}
\IEEEauthorblockA{\textit{Computational Science and Eng.} \\
\textit{Georgia Institute of Technology}\\
Atlanta, Georgia}\\
\IEEEauthorblockN{Spencer H. Bryngelson}
\IEEEauthorblockA{\textit{Computational Science and Eng.} \\
\textit{Georgia Institute of Technology}\\
Atlanta, Georgia \\
\href{mailto:shb@gatech.edu}{shb@gatech.edu}
}
}

\maketitle

\begin{abstract}
GPUs are the heart of the latest generations of supercomputers. 
We efficiently accelerate a compressible multiphase flow solver via OpenACC on NVIDIA and AMD Instinct GPUs. 
Optimization is accomplished by specifying the directive clauses \texttt{gang vector} and \texttt{collapse}.
Further speedups of six and ten times are achieved by packing user-defined types into coalesced multidimensional arrays and manual inlining via metaprogramming. 
Additional optimizations yield seven-times speedup of array packing and thirty-times speedup of select kernels on Frontier.
Weak scaling efficiencies of 97\% and 95\% are observed when scaling to 50\% of Summit and 87\% of Frontier.
Strong scaling efficiencies of 84\% and 81\% are observed when increasing the device count by a factor of 8 and 16 on V100 and MI250X hardware. 
The strong scaling efficiency of AMD’s MI250X increases to 92\% when increasing the device count by a factor of 16 when GPU-aware MPI is used for communication.
\end{abstract}

\begin{IEEEkeywords}
OpenACC, directive offloading, NVIDIA GPU, AMD GPU
\end{IEEEkeywords}

\section{Introduction}
GPUs are important in increasing the computational power of the newest and fastest supercomputers.
OLCF Titan was the first leadership\Hyphdash*class supercomputer that used GPUs for scientific computation.
Titan was followed by the V100 generation of NVIDIA machines in the United States, like OLCF's Summit and LLNL's Sierra, which both spent time as the fastest supercomputer in the world.
These V100-based machines have been replaced with AMD-powered exascale machines, OCLF's Frontier, and LLNL's El Capitan.
At the same time, the NVIDIA-powered machine JUPITER, an exascale machine at the J\"{u}lich Super Computing Center in Europe, is being commissioned.
Conducting simulations with hundreds of billions of grid cells on leadership\Hyphdash*class supercomputers powered by multiple hardware vendors requires using GPUs with efficiency and portability.

Several teams have made porting efforts to use the AMD hardware featured on Frontier.
URANOS-2.0~\cite{uranos20} is a Fortran-based compressible Navier--Stokes solver that supports NVIDIA and AMD GPUs with OpenACC offloading.
STREAmS-2~\cite{streams2} is also a compressible Navier--Stokes solver that supports NVIDIA and AMD~GPUs.
STREAMs supports NVIDIA GPUs using CUDA Fortran and AMD GPUs using HIPFort.
An OpenMP port of STREAMs adds support for Intel GPUs with directive-based offloading~\cite{streamsOMP}.

MFC~\cite{bryngelson2021}, discussed herein, solves compressible multiphase flow problems with Fortran and supports NVIDIA and AMD offloading via OpenACC~\cite{OpenACC}.
OpenACC requires the user to identify regions in code that can benefit from hardware acceleration and indicate to the compiler that it should create hardware-accelerated kernels for these regions.
The compiler then generates optimized kernels for the identified regions.
NVIDIA's NVHPC supports OpenACC offloading on NVIDIA GPUs, and HPE's Cray Compiler Environment (CCE) supports OpenACC offloading on NVIDIA and AMD GPUs.
GNU~14+ and Flang~18+ also support OpenACC, though their relative OpenACC immaturity prevents their use in our application.
Due to the directive-based implementation, the resulting implementation is also performant on CPU architectures.
Our implementation is compatible with ARM64 and x86\Hyphdash*64 CPU architectures when compiled without OpenACC.

We evaluate our implementation's roofline and scaling performance on leadership-class NVIDIA and AMD supercomputers.
Summit is an NVIDIA V100 machine with 27,648 GPUs and achieves a peak of $\SI{148.6}{\peta\flop\per\second}$ on the LINPACK benchmark.
Frontier is an AMD~MI250X-based supercomputer with 37632 GPUs and achieves a peak of $\SI{1.2}{\exa\flop\per\second}$ on the LINPACK benchmark, making it eight times faster than Summit.
Each AMD~MI250X on Frontier contains two graphics compute dies (GCDs), which are effectively separate hardware devices.
The newer NVIDIA A100, H100, and GH200 devices, as well as the V100 and MI250X, are used to compare grind time and its makeup.

We introduce the physical model and numerical method for simulating compressible multiphase flows in \cref{sec:background}.
The background is followed by details regarding the implementation and optimizations that yield the performance and portability observed on NVIDIA and AMD GPUs in \cref{sec:implementation}.
Next, we provide detailed performance results supporting our claims of performance and portability in \cref{sec:performance}.
We close by demonstrating the simulation capabilities via three fluid flow demonstration cases simulated on three different GPUs in \cref{sec:examples}.

\section{Background}\label{sec:background}
\subsection{Physical Model}
\label{s:model}
We use Baer--Nunziato~\cite{andrianov2004} type models for multiphase flows with a diffuse interface capturing scheme.
Diffuse interface schemes allow for artificial diffusion between phases and avoid the need for mesh management and unique treatments to ensure conservation required by interface tracking schemes~\cite{Saurel2018}.
The original model of Baer and Nunziato allows for disequilibrium in pressure and velocity between fluid phases.
We use the reduced 5-equation model for multiphase flows of~\citet{allaire2002}.
The Allaire model enforces pressure and velocity equilibrium between phases and is suitable for many multiphase flow problems.
For two components, the Allaire model is given by
\begin{align*}
    \pdiff[\alpha_i]{t} + \nabla \cdot (\alpha_i\rho_i\bu) &= 0, \\
    \pdiff[\rho \bu]{t} + \nabla \cdot (\bu \otimes \bu + p\bI) &= 0, \\
    \pdiff[\rho E]{t} + \nabla \cdot \left[ (\rho E + p)\bu \right] &= 0, \\
    \pdiff[\alpha_i]{t} + \bu \cdot \nabla \alpha_i &= 0,
\end{align*}
where $\rho$, $\bu$, $p$, and $E$ are the mixture density, velocity, pressure, and energy, and $\alpha_i$ are the volume fractions of component $i$.
The system of equations is closed using the stiffened gas equation of state (EOS), which faithfully models many liquids and gases~\cite{Metayer2016}:
\begin{equation*}
    \rho E = \frac{1}{\gamma - 1}p + \frac{\gamma \pi_\infty}{\gamma - 1}.
\end{equation*}
The parameters $\gamma$ and $\pi_\infty$ are the mixture ratio of specific heats and liquid stiffness.
The liquid stiffness parameter allows for liquids to be modeled as if they were high-pressure gasses.

\subsection{Numerical Method}
A finite volume scheme that follows Coralic and Colonius~\cite{Coralic2014} solves the model of~\cref{s:model}.
These methods are implemented in MFC~\cite{bryngelson2021}, a GPU-accelerated compressible CFD solver~\cite{elwasif2023application, radhakrishnan22, radhakrishnan24} written in Fortran (2003).
The governing equations are discretized on a structured grid as
\begin{equation}
    \pdiff[\bq]{t} + \pdiff[\bF^x(\bq)]{x}
        + \pdiff[\bF^y(\bq)]{y} +
        \pdiff[\bF^z(\bq)]{z} + 
        \bm{h}(\bq) \nabla \cdot \bu
        = 0
    \label{eqn:discretization}
\end{equation}
where $\bq$ and $\bF$ are the vectors of conservative variables and fluxes and $\bh(\bq)$ is a volume averaged source term.
\Cref{eqn:discretization} is integrated in space across each cell center as
\begin{equation*}
\begin{aligned}
    \pdiff[\bq_{m,n,p}]{t} &= \frac{1}{\Delta x_j}\left(\bF^x_{m - 1/2,n,p} - \bF^x_{m + 1/2,n,p}\right) \\
    & + \frac{1}{\Delta y_j}\left(\bF^y_{m, n- 1/2,p} - \bF^y_{m,n+1/2,p}\right) \\
    & + \frac{1}{\Delta z_k}\left(\bF^z_{m,n,p- 1/2} - \bF^z_{m,n,p+ 1/2}\right) \\
    & - \bm{h}(\bq_{m,n,p})\left(\nabla\cdot \bu\right)_{m,n,p} \label{eqn:rhs}.
\end{aligned}
\end{equation*}
The flux terms $\bF$ are obtained by averaging over cell interfaces.
The fluxes at cell interfaces are computed via the HLLC approximate Riemann solver~\cite{toro2009riemann} using third or fifth-order accurate WENO reconstructions~\cite{Shu2006}.

\section{Implementation}\label{sec:implementation}
\subsection{Domain Decomposition and I/O}

The computational domain is decomposed into 3D blocks across processors to exploit distributed computing capabilities.
Blocks with uniform dimensions in each direction are used rather than splitting in one direction (slabs) or two directions (pencils).
Compressible flow solvers do not require global communication.
Blocks reduce the overall communication cost by minimizing the surface-to-volume ratio of each process's domain.
Minimizing the surface-to-volume ratio keeps the buffer regions for nearest-neighbor communication between processes small relative to each process's domain. 

The domain is discretized using a structured mesh.
Local mesh refinement is implemented using a hyperbolic tangent function \cite{VINOKUR1983}.
Cartesian, axisymmetric, and cylindrical coordinates are supported.
3D cylindrical grids make use of \texttt{FFTW}~\cite{FFTW} on CPUs, \texttt{cuFFT}~\cite{cuFFT} on NVIDIA GPUs, and \texttt{hipFFT}~\cite{hipFFT} on AMD GPUs to apply a low-pass filter to the flow variables in the azimuthal direction.
Low-pass filtering removes the high-frequency content, alleviating the restrictive CFL condition from the grid cells near the axis.

The third- and fifth-order accurate reconstructions that improve asymptotic convergence rates require information at the solution state outside each processor's local domain.
A halo exchange between adjacent processors is required in each dimension at each time step.
Each process packs its corresponding buffer region into a 1D array for compatibility with MPI subroutines.
Each process performs an \texttt{MPI\_sendrecv} with relevant neighbor processes in the current dimension and unpacks the received buffer.
This halo exchange process gives each process the information necessary to perform third- and fifth-order accurate WENO reconstructions.

Unsteady compressible flow simulations require I/O operations at intervals of $\mathcal{O}(10^3)$ time steps.
I/O is performed using MPI I/O via collective operations into a shared binary file or within each process, each writing a binary file.
Before Frontier, MFC relied on the shared binary file approach.
When scaling to 65,536 GCDs on Frontier, we witnessed increased I/O times when creating MPI I/O shared binary files and found a file-per-process approach more appropriate.
Write access is allowed in waves of 128 processes, with each wave offset by a set number of double-precision multiplication operations.
Allowing file access in small waves avoids overwhelming the file system with metadata creation.

Host code reads the MPI I/O binary files and creates SILO~\cite{millersilo} files that can be visualized with several free tools, including Paraview~\cite{Ahrens2005ParaViewAE} and VisIt~\cite{VisIt}.

\subsection{GPU Offloading}

We use OpenACC \cite{wienke2012} with NVHPC and CCE compilers for directive-based offloading of all compute kernels on NVIDIA and AMD accelerators.
Once preprocessing has been performed on CPUs, the initial state variables are transferred to the GPUs, and all subsequent computation is performed on the GPU.
The CPU facilitates local MPI communication of relatively small buffer regions (if GPU-aware MPI is unavailable) and performs I/O operations once the initial state has been transferred to the accelerators.
The relatively expensive GPU--CPU data transfer required for I/O occurs at intervals of $\mathcal{O}(10^3)$ time steps, making it negligible to the overall runtime.
Directive-based offloading with OpenACC requires only identifying independent loops and specifying the desired level of parallelism.
Once independent loops are identified, the compiler generates optimized kernels. 
We use \texttt{cuTENSOR}~\cite{cuTENSOR} and \texttt{cuFFT} libraries on NVIDIA hardware and \texttt{hipBLAS}~\cite{hipBLAS} and \texttt{hipFFT} libraries on AMD hardware for optimized array packing and Fourier transforms.

\subsection{Optimization}

\begin{lstlisting}[caption={Directive setup for a core OpenACC kernel},label={lst:kernel},xleftmargin = 0mm,float,escapeinside={(*}{*)}]
!$acc parallel loop vector gang collapse(3) &
!$acc default(present) private(...)
do l = is3%beg, is3%end         !coordinate 3
    do k = is2%beg, is2%end     !coordinate 2
        do j = is1%beg, is1%end !coordinate 1
            !$acc loop seq
            do i = 1, num_PDEs
                !!> Core kernel,
                !!> O(100) arithmetic operations
            end do
        end do
    end do
end do
!$acc end parallel loop
\end{lstlisting}

\Cref{lst:kernel} shows the directives used for a kernel in which $\mathcal{O}(10^2)$ computations must be performed for each equation in each cell.
Loops \texttt{j}, \texttt{k}, and \texttt{l} loop over $\mathcal{O}(10^2)$ elements each and loop \texttt{i} loops over $\mathcal{O}(1)$ elements for common two-phase flow problems like shock droplet and shock bubble interaction.
The two most expensive kernels, approximate Riemann solve and WENO reconstruction, have this loop structure.

OpenACC kernels distribute the workload using gangs, workers, and vectors corresponding to blocks, warps, and threads in CUDA notation.
The default behavior of the \texttt{parallel loop} directive in OpenACC is to split loop iterations across gangs, leading to each block using a single OpenACC vector and under utilizing resources.
We append \texttt{gang vector} clauses to all \texttt{parallel loop} directives to split the loop iterations across multiple gangs with fixed vector length \cite{sunita2017}.
Further optimization is obtained by collapsing the \texttt{j}, \texttt{k}, and \texttt{l} loops in \cref{lst:kernel} into a single loop using \texttt{collapse(3)}, which allows the compiler to select the optimal gang and vector size for a given problem and architecture.
We find that the innermost loop over \texttt{i} in \cref{lst:kernel} benefits from serialization via \texttt{!\$acc loop seq} due to its small loop range of $\mathcal{O}(1)$ for general two-phase problems.

Additional optimization in the most expensive kernels is achieved by packing user-defined struct-of-array data types into flattened multidimensional arrays.
The compiler can perform aggressive optimizations on flattened multidimensional arrays that are not possible with user-defined types.
Using multidimensional arrays rather than user-defined types for a representative two-phase problem with one million grid cells, a sixfold speedup in the WENO kernel was observed. 

Additional speedup is made possible by reshaping arrays for coalesced memory access in the most expensive kernels.
Memory coalescence allows for increased throughput of high bandwidth memory on accelerators.
Coalescing memory results in a ten-times speedup in the WENO kernel for a representative two-phase flow problem with one million grid cells. 
This reduction outweighs the cost required to transpose the arrays.
With NVHPC and NVIDIA hardware, the \texttt{cuTENSOR} library performed transposes with similar performance to fully collapsed OpenACC loops.
With CCE and MI250X hardware, the \texttt{hipBLAS} library showed a seven-times speedup over fully collapsed OpenACC loops for the same transpose operations.

Metaprogramming, enabled by Fypp~\cite{fypp}, further improves GPU kernel performance. 
Sometimes, the compiler does not automatically inline serial subroutines within GPU kernels across modules.
Using Fypp allows these subroutines to be inlined by programmer directives for compiler optimization.
Fypp does not generate any code that could not be written manually.
However, it does generate code that would be tedious to write manually, reducing overall line count and improving code readability.
Inlining serial subroutines via programmer directives with Fypp prevents a tenfold slowdown of the Riemann and WENO kernels that would otherwise call serial subroutines.

\subsection{Frontier and CCE Specific Optimization}\label{sec:frontierOpt}

Profiles showed that the MI250X spent a significant amount of its runtime packing struct-of-array data into 4D arrays to facilitate lowest-rank coalesced memory access in the destination array for the approximate Riemann solver and WENO kernels.
The fully collapsed three- or four-loop OpenACC kernels that perform well on NVIDIA's hardware with NVHPC execute slowly on the MI250X with CCE.
We suspect that the slowdown is due to the $\SI{8}{\mega\byte}$ L2 cache of the MI250X, though it could also be a result of poor optimizations by the compiler.

The best practice to improve matrix transpose efficiency is to use the BLAS extension GEAM.
The ROCm software stack provides the \texttt{hipBLAS} library, which contains a strided, batched, double-precision GEAM operation that efficiently transposes the first two indices of an array.
Arbitrary index transposes can be performed using index fusing and batched transposes.
A seven-fold reduction in computational time is achieved for these kernels when using \texttt{hipBLAS} libraries.

The CCE Fortran compiler with OpenACC offloading performed poorly when encountering any variable in a \texttt{private} clause with an unknown compile-time size.
In this case, the compiler allocates memory for that variable at run time after the thread~block starts and the size is known.
Device-side allocations are expensive on current AMD~GPUs because they require the kernel to write information into a special buffer, the host to read this buffer, perform an action, and inform the kernel it can continue.
One kernel in which this phenomenon was particularly problematic went from taking 90\% of the total runtime to just 3\% of the total runtime when just one, $\mathcal{O}(1)$-element array in its \texttt{private} clause had its size declared at compile time.

\subsection{Library Offload Implementations}

Offloading suitable computations to vendor-provided libraries provides access to hardware-optimized implementations without manual tuning.
We utilize vendor-provided libraries to perform array transposes and calculate fast Fourier transforms.
\Cref{lst:SF} defines the \texttt{scalar\_field} data type that is necessary to understand the array transpose implementations.

\begin{lstlisting}[caption={Scalar Field Derived Type},label={lst:SF},xleftmargin = 0mm,float]
type scalar_field
    real(kind(0d0)), pointer, &
        dimension(:, :, :) :: sf => null()
end type scalar_field
\end{lstlisting}

\Cref{lst:transposeNV} shows how \texttt{cuTENSOR} is used to convert an array of scalar fields into a flattened 4D array with coalesced memory at each initialization of the WENO reconstruction step.
In the NVIDIA implementation, the array of scalar fields is first packed into a 4D temporary array using fully collapsed OpenACC loops.
The temporary array \texttt{v\_temp} is generated only once per initialization of the WENO reconstruction step when coalescing memory in the $x$-direction and reused in the $y$- and $z$-directions.
Calls to \texttt{reshape} bracketed by \texttt{!\$acc host\_data use\_device()} directives indicate that \texttt{cuTENSOR} should be used to perform the specified index shifts on the GPU.

\begin{lstlisting}[caption={GEAM transpose with \texttt{cuTENSOR} from index order (1,2,3,4) to (3,2,1,4)},label={lst:transposeNV},xleftmargin = 0mm,float=*,floatplacement=tbp]
subroutine s_GEAM_transpose(v_vf, v_sf_t, n1, n2, n3, n4)
    use CuTensorEx
    type(scalar_field), dimension(:) :: v_vf ! Initial array of scalar fields
    real(kind(0d0)), dimension(:,:,:,:) :: v_sf_t, v_temp ! Transposed and temporary array
    integer :: n1, n2, n3, n4 ! Array dimensions

    !$acc parallel loop collapse(4) gang vector default(present)
    do j = 1, n4
        do q = 1, n3
            do l = 1, n2
                do k = 1, n1
                    v_temp(k, l, q, j) = v_vf(j)%sf(k, l, q)
                end do
            end do
        end do
    end do
    !$acc end parallel loop

    !$acc host_data use_device(v_temp, v_sf_t)
    v_sf_t = reshape(v_temp, shape=[n3, n2, n1, n4], order=[3, 2, 1, 4])
    !$acc end host_data
end subroutine s_GEAM_transpose
\end{lstlisting}

\Cref{lst:transposeAMD} shows the more involved approach of packing an array of scalar fields into a flattened 4D array with coalesced memory using \texttt{hipBLAS}.
We include only how the $z$-direction memory coalescence is achieved for brevity.
Coalescence in the $x$- and $y$-direction are completed via fully collapsed OpenACC loops and a single strided and batched \texttt{hipBLAS} GEAM call.
Memory coalescence in the $z$-direction requires two GEAM operations because it swaps the first and third indices.
The first call is to a strided, batched GEAM operation and swaps the first and second indices ($\bA_{ijk}\rightarrow\bA_{jik}$).
A strided, batched GEAM can be used for this operation because it is equivalent to $k$ permutations of $\bA_{ij}$ to $\bA_{ji}$.
The second GEAM operation groups the $i$ and $j$ indices and performs the permutation $\bA_{(ji)k}$ to $\bA_{k(ji)}$ using an unbatched GEAM operation.
An unbatched operation is used because this is a single permutation between a $j\times i$ dimension tensor with a $k$ dimension tensor.

\begin{lstlisting}[caption={GEAM transpose with \texttt{hipBLAS} from index order (1,2,3,4) to (3,2,1,4)},label={lst:transposeAMD},xleftmargin = 0mm,float=*,floatplacement=tbp]
subroutine s_GEAM_transpose(v_vf, v_sf_t, n1, n2, n3, n4)
    use hipfort
    use hipfort_hipblas
    use hipfort_check
    
    type(scalar_field), dimension(:) :: v_vf ! Initial array of scalar fields
    real(kind(0d0)), dimension(:,:,:,:) :: v_sf_t, transpose_tmp ! Transposed and temporary array
    integer :: n1, n2, n3, n4 ! Array dimensions
    integer :: j ! Loop iterator

    for j = 1, n4
        !$acc host_data use_device(v_vf(j), v_sf_t, transpose_tmp)
        call hipblascheck(hipblasdgeamstridedbatched(handle, HIPBLAS_OP_T, HIPBLAS_OP_T, n1, n2, &
            1.0_8, c_loc(v_vf(j)), n2,int(n1*n2,c_int64_t),     &
            0.0_8, c_loc(v_vf(j)), n2,int(n1*n2,c_int64_t),     &
            c_loc(transpose_tmp, n1,int(n1*n2,c_int64_t),n3))

        call hipblascheck(hipblasdgeam(handle, HIPBLAS_OP_T, HIPBLAS_OP_T,  n1, n2*n3, &
            1.0_8, c_loc(transpose_tmp), n2*n3,&
            0.0_8, c_loc(transpose_tmp), n2*n3, &
            c_loc(v_sf_t(:,:,:,j)), n1))
        !$acc end host_data
        call hipCheck(hipDeviceSynchronize())
    end do
end subroutine s_GEAM_transpose
\end{lstlisting}

Library calls to \texttt{cuFFT} and \texttt{hipFFT} are more straightforward.
\Cref{lst:fftNV} and \cref{lst:fftAMD} demonstrate how \mbox{OpenACC} directives are used with \texttt{cuFFT} and \texttt{hipFFT} to compute the forward Fourier transform.
Inverse Fourier transforms are done by replacing \texttt{cufftExecD2Z} and \texttt{hipfftExecD2Z} with \texttt{cufftExecZ2D} and \texttt{hipfftExecZ2D} using the same OpenACC directives.

\begin{lstlisting}[caption={Fast Fourier transform with \texttt{cuFFT}.},label={lst:fftNV},xleftmargin = 0mm,float]
!$acc host_data use_device(data_real,data_cmplx)
ierr = cufftExecD2Z(fwd_plan, data_real, data_cmplx)
!$acc end host_data
\end{lstlisting}

\begin{lstlisting}[caption={Fast Fourier transform with \texttt{hipFFT}.},label={lst:fftAMD},xleftmargin = 0mm,float]
!$acc host_data use_device(data_real,data_cmplx)
ierr = hipfftExecD2Z(fwd_plan, c_loc(data_real), c_loc(data_cmplx))
call hipCheck(hipDeviceSynchronize())
!$acc end host_data
\end{lstlisting}

\subsection{Validation}

MFC has been validated against experimental results for several canonical problems in multiphase fluid dynamics, such as shock bubble and shock droplet interaction, spherical bubble collapse, and Taylor--Green vortices~\cite{bryngelson2021}.
Every MFC pull request comprises over 250 test cases using NVIDIA, GCC, CCE, and Intel compilers on CPUs, AMD GPUs, and NVIDIA GPUs.
This rigorous testing helps ensure that simulation results remain independent of compiler or hardware.

\section{Performance}\label{sec:performance}
We begin our performance summary by presenting results for MFC on Summit (NVIDIA~V100) and Frontier (AMD~MI250X).
More specifically, we show kernel-level performance of the most expensive kernels, weak scaling results up to 50\% of OLCF Summit and 87\% of OLCF Frontier, and strong scaling results over a 16-times and 512-times increase in device count on OLCF Summit and OLCF Frontier.
We then show speedups over the fastest tested CPUs from AMD, Intel, NVIDIA, and IBM at the time of writing.
After this, we detail the breakdown of time spent in the most expensive kernels and time spent packing arrays on NVIDIA's most recent hardware and compare the results with AMD's MI250X.
We find that memory bandwidth and L2 cache play a role in a 3.71- and 2.62-times increase array packing runtime on V100 and MI250X GPUs compared to an A100 GPU.

\subsection{Kernel-Level Performance}

\begin{figure}
    \centering
    \includegraphics{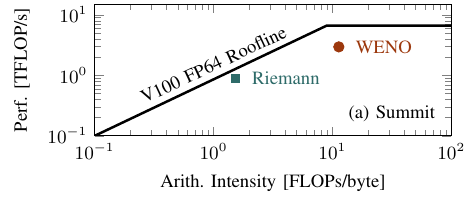}
    \includegraphics{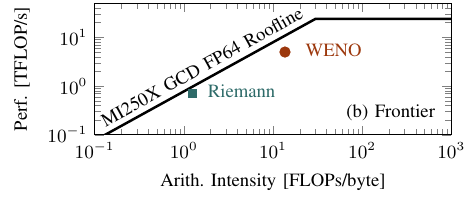}
    \caption{
        Roofline performance of the most expensive kernels on (a) OLCF Summit and (b) OLCF Frontier.
        The Riemann solver and WENO kernels use 13\% and 45\% of peak FLOPS on OLCF Summit.
        On OLCF Frontier, the Riemann solver and WENO kernels use 3\% and 21\% of peak FLOPS.
    }
    \label{fig:roof}
\end{figure}

Roofline performance is measured using NVIDIA's \texttt{nsys-compute} on NVIDIA hardware and CCE's \texttt{omniperf} on AMD hardware.
\Cref{fig:roof} shows the roofline performance of the approximate Riemann solve and WENO reconstruction kernels on V100 and MI250x GPUs.
These two kernels are shown because they are the two most expensive kernels, accounting for 63\% and 56\% of the grind time spent on compute-focused tasks while only making up 10\% of the total number of kernels on the V100 and MI250X.
On the V100, the approximate Riemann solve is memory\Hyphdash*bound, while the WENO reconstruction is compute\Hyphdash*bound.
On the MI250X, both kernels are memory\Hyphdash*bound because its transition from memory to compute bound occurs at an arithmetic intensity 3.4 times that of a V100.
The WENO kernel achieves 45\% of V100 peak FP64 FLOPS and 21\% of MI250X peak FP64 FLOPS.
The memory\Hyphdash*bound approximate Riemann solve kernel achieves 13\% of V100 peak FP64 FLOPS and 3\% of MI250X peak FP64 FLOPS.
The decreased roofline performance of the MI250X is likely due to its $\SI{8}{\mega\byte}$ L2.

\subsection{Weak Scaling Performance}

\begin{figure}
    \centering
    \includegraphics{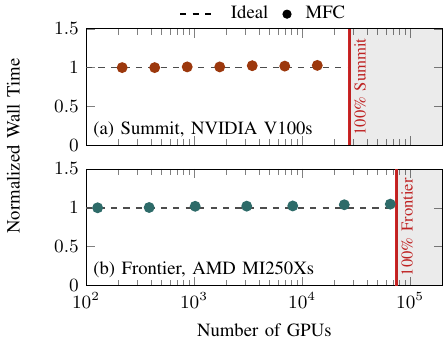}
    \caption{
    Weak scaling results on (a) OLCF Summit and (b) OLCF Frontier.
    On OLCF Summit, MFC scales from 128 V100 GPUs to 13824  V100 GPUs (50\% of the machine) with 97\% efficiency.
    On OLCF Frontier, MFC scales from 128 MI250X GCDs to 65536 MI250X GCDs (87\% of the machine) with 95\% efficiency.}
    \label{fig:ws}
\end{figure}

Weak scaling performance is measured using the wall runtime while increasing the problem size with the number of devices so that each device maintains the same amount of work.
\Cref{fig:ws} shows weak scaling results for a representative two-phase problem on OLCF Summit and OLCF Frontier.
MFC scales from 128 to 13825 V100 GPUs (50\% of the machine) on OLCF Summit with 97\% efficiency and from 128 to 65536 MI250X GCDs (87\% of the machine) on OLCF Frontier with 95\% efficiency.
The wall times are normalized by the wall time of the base case, which has 128 V100 GPUs on Summit and 128 MI250X GCDs on Frontier.
These results are unsurprising because the nearest\Hyphdash*neighbor communication required to send and receive buffer regions remains constant as the number of processes increases while the grid cells per process remain constant, and no significant collective communication is required.

\subsection{Strong Scaling Performance}

\begin{figure}
    \centering
    \includegraphics{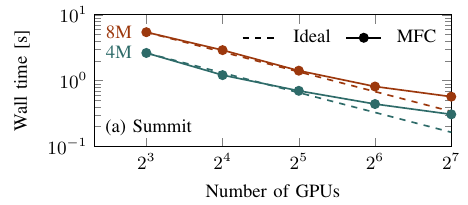}
    \includegraphics{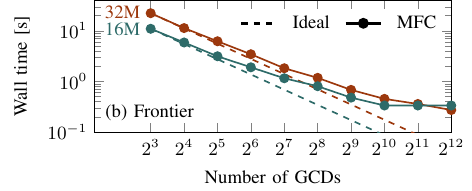}
    \caption{Strong scaling results on (a) OLCF Summit and (b) OLCF Frontier.
    On OLCF Summit, a simulation with 8 million cells per GPU maintains 84\% of ideal performance when increasing the GPU count by a factor of 8.
    On OLCF Frontier, a simulation with 32 million cells per GCD maintains 81\% of ideal performance when increasing the GCD count by a factor of 16.}
    \label{fig:ss}
\end{figure}

Strong scaling performance is measured using the wall runtime and increasing the number of devices while maintaining a constant problem size.
\Cref{fig:ss} shows the strong calling performance of MFC on OLCF~Summit and OLCF~Frontier for the same representative two-phase problem used for weak scaling tests.
A V100 GPU maintains 84\% of ideal performance when the GPU count is increased by a factor of eight for a two-phase problem with 8 million cells per GPU.
An MI250X GCD maintains 81\% of ideal performance when the GCD count is increased by a factor of 16 for a two-fluid problem with 32 million cells per GCD.
A smaller problem with half as many grid cells per device scales with poorer efficiency due to the increased MPI costs relative to compute costs.
This eventually leads to the flatline observed in the 16 million cell results on OLCF Frontier in \cref{fig:ss}.

\begin{figure}
    \centering
    \includegraphics{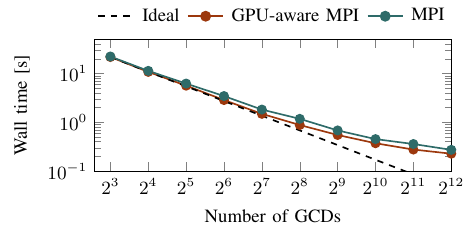}
    \caption{Strong scaling on OLCF Frontier with and without GPU-aware MPI.
    With GPU-aware MPI, a simulation with 32 million cells per GCD maintains 92\% of ideal performance when increasing the GCD count by a factor of 16.
    This is a 14\% increase over the 81\% of ideal performance maintained when increasing device count by a factor of 16 without GPU-aware MPI.}
    \label{fig:ssRDMA}
\end{figure}

\Cref{fig:ssRDMA} shows the improvement in strong scaling performance achieved by using GPU-aware MPI on Frontier with HIP-coupled MPI libraries. 
With GPU-aware MPI enabled, a simulation with 32 million cells per MI250X GCD maintains 92\% of ideal performance when the GCD count is increased by a factor of 16.
This is a 14\% increase from the 81\% of ideal performance maintained when increasing the device count by a factor 16 without GPU-aware MPI for the same problem size.

\subsection{Speedup}

\begin{figure}
    \centering
    \includegraphics{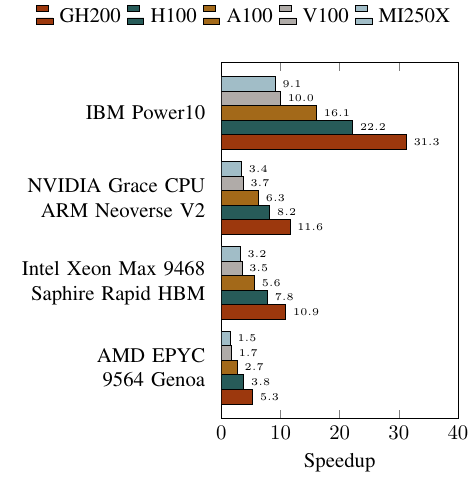}
    \caption{Relative speed up in grind time for the optimized code executed on GPUs, as labeled, compared to the fastest benchmarked CPUs (all cores used) offered by AMD, NVIDIA, Intel, and IBM.
    Larger numbers indicate a large speedup of the single GPU over the single CPU socket.}
    \label{fig:speedup}
\end{figure}

\Cref{fig:speedup} shows the speed up of the GH200, H100, A100 and V100 GPUs and MI250X GCD over the fastest benchmarked CPUs from AMD, NVIDIA, Intel, and IBM.
Speedup is measured using the grind time in nanoseconds per grid cell, PDE, and right-hand-side evaluation, with CPU runtimes being normalized by the total number of cores.
Normalization by the number of cores means the results should be interpreted as one GPU die is \textit{X}-times faster than one CPU die.
At the time of writing, the AMD EPYC 9564 Genoa is the fastest tested CPU, with the tested GPUs achieving speedups of only 1.5 to 5.3 times.
Intel's Xeon Max 9468 Saphire Rapid HBM and NVIDIA's ARM Neoverse V2 Grace CPU perform similarly, with the tested GPUs achieving between 3 and 11 times speedup.
The older IBM Power 10 is slower than the newer chips from NVIDIA, Intel, and AMD, with the tested GPUs achieving 9.1 to 31.3 times speedups.

\section{Compute Breakdown}

In the following section, NVHPC HPC SDK 22.11 is used to collect results for V100 and A100 devices, NVHPC HPC SDK 24.5 is used to collect results for H100 and GH200 hardware, and HPE Cray Compiling Environment (CCE) 16.0.1 is for the AMD MI250X.
NVIDIA's \texttt{nsight-compute} and CCE's \texttt{roc-prof} are used on NVIDIA and AMD devices for measuring kernel run times.

\begin{figure}
    \centering
    \includegraphics{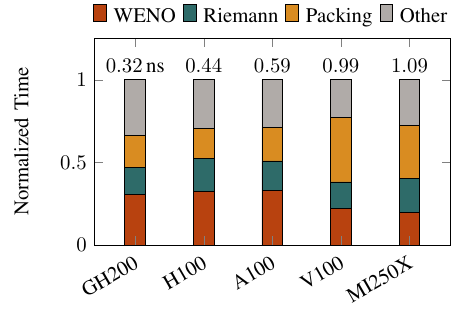}
    \caption{The percentage of normalized runtime spent in the most expensive kernels and array packing for an example problem with 8 million grid cells per device on five GPU devices: (1) NV GH200, (2) NV H100 SXM, (3) NV A100 PCIe, (4) NV V100 PCIe, and (5) AMD MI250X GCD.
    Numbers at the top of each column indicate the simulation grind time in nanoseconds per grid cell, PDE, and right-hand-side evaluation}
    \label{fig:bdNorm}
\end{figure}

\Cref{fig:bdNorm} shows the percentage of time spent in each of the most expensive kernels and array packing for a two\Hyphdash*phase problem with 8 million grid cells in 3D.
NVIDIA's most recent GPUs at the time of writing, the GH200, H100, and A100, spend a similar percentage of simulation time in each kernel.
However, the V100 and MI250X GCD spend a more significant percentage of runtime packing arrays.
On the V100, slow array packing results from the low memory bandwidth of $\SI{900}{\giga\byte\per\second}$.
We suspect that slow array packing on the MI250X results from the $\SI{8}{\mega\byte}$ L2 cache, which leads to a high rate of L2 cache misses.
Kernel-level profiles of array packing routines show that the MI250X has three times the L2 cache misses of an A100.
NVIDIA A100, H100, and GH200 have memory bandwidths of $\SI{2}{\tera\byte\per\second}$, $\SI{3.35}{\tera\byte\per\second}$, and $\SI{4}{\tera\byte\per\second}$ and L2 cache sizes of $\SI{40}{\mega\byte},$ $\SI{50}{\mega\byte},$ and $\SI{50}{\mega\byte}$.
The higher memory bandwidth and larger L2 cache both positively impact packing performance.

\begin{figure}
    \centering
    \includegraphics{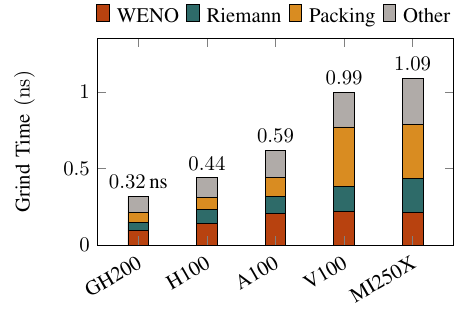}
    \caption{Grind time spent in the most expensive kernels and array packing for an example problem with 8 million grid cells per device on five GPU devices: (1) NV GH200, (2) NV H100 SXM, (3) NV A100 PCIe, (4) NV V100 PCIe, (5) AMD MI250X GCD.
    The grind time is measured in nanoseconds per grid cell, PDE, and right-hand-side evaluation.
    }
    \label{fig:bdAbs}
\end{figure}

\Cref{fig:bdAbs} shows the time spent in each of the most expensive kernels and transposing arrays for a two-phase example problem with 8 million grid cells in 3D.
The consequences of memory availability are more pronounced when grind times are shown than normalized times.
The NVIDIA V100 and AMD MI250X GCD take 5\% and 4.5\% longer to perform the compute-bound WENO kernel than an A100.
The increased runtime on the V100 results from it having 72\% of the peak FLOPS of an A100.
On the MI250X, the increased runtime likely results from the L2 cache being one-fifth the size of an A100.
The MI250X makes up for its $\SI{8}{\mega\byte}$ cache by having 2.5 times the peak FLOPS of an A100.
The memory-bound Riemann kernel takes 48\% and 103\% longer on a V100 and MI250X than an A100.
The most significant slowdown results from the time spent packing arrays by the V100 and MI250X.
For the same problem size, the V100 and MI250X take 3.71 and 2.62 times longer to pack arrays than an A100.
From this, we conclude that optimizing data movement is an important factor in the overall performance of GPUs.

\section{Example Simulations}\label{sec:examples}
This section shows example simulations run on NVIDIA and AMD GPUs, demonstrating our implementation.

\subsection{Shock Droplet}

\begin{figure}
    \centering
    \includegraphics[width=\columnwidth,clip,trim={0.2in 0.8in 3in 1in}]{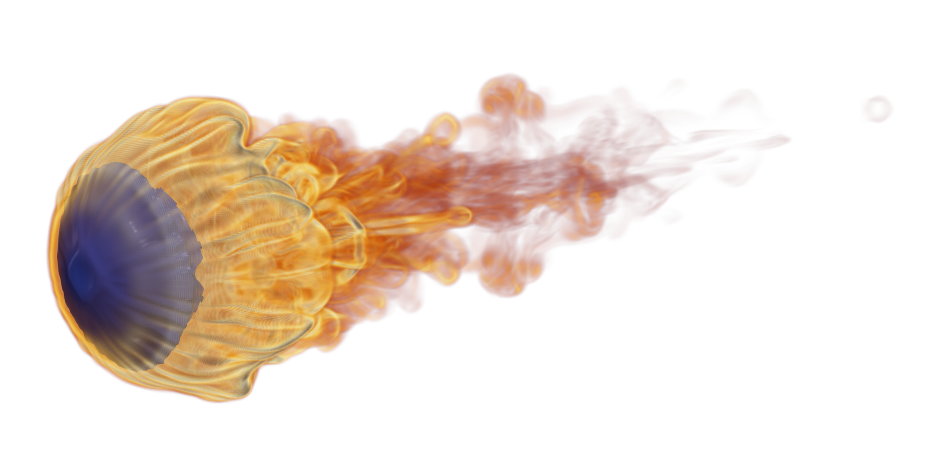}
    \caption{
        Vorticity around a shock droplet in air.
        The droplet's surface is blue, and the vorticity magnitude is shown in the orange and red volume rendering around and behind the droplet.
        Flow is from left to right, and darker colors correspond to higher vorticity magnitudes.
    }
    \label{fig:sd}
\end{figure}

The first example case shows a shock droplet interaction.
More specifically, a Mach~1.46 air shock impinging a water droplet.
This simulation domain was discretized into 2~billion grid~cells and advanced through 100~thousand time steps using 960~V100 GPUs in 2~hours on OLCF Summit.
\Cref{fig:sd} shows the droplet surface in blue and a volume rendering of vorticity behind the droplet in orange.

\subsection{Flow over an airfoil}

\begin{figure}
    \centering
    \includegraphics[width=\columnwidth,clip,trim={0 0 1.2in 0.9in}]{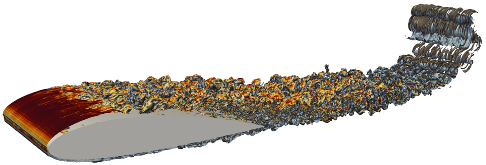}
    \caption{Vorticity shedding of a NACA 2412 airfoil with a 15~degree angle of attack.
    The airfoil is gray, and the vorticity contour is colored by the span-wise vorticity, increasing in magnitude from red to gray to blue.
    }
    \label{fig:airfoil}
\end{figure}

The next example case shows the flow of air over a NACA~2412 airfoil modeled using the ghost cell immersed boundary method.
This simulation resolves 500~grid~cells in the chord length of the airfoil and comprises a total of 2.25~billion grid~cells.
It was advanced through 93 thousand time steps using 128~A100 GPUs in 19~hours on NCSA~Delta.
\Cref{fig:airfoil} shows the vortex shedding around the airfoil colored by the span-wise vorticity.

\subsection{Shock bubble cloud}

\begin{figure}
    \centering
    \begin{tikzpicture}
        \node (image) at (0,0) {\includegraphics[width=1in]{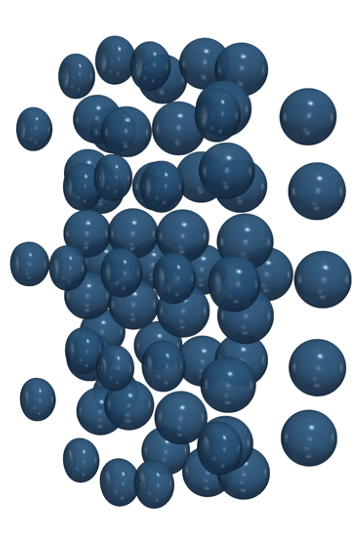}}; 
        \node (image) at (1in,0) {\includegraphics[width=1in]{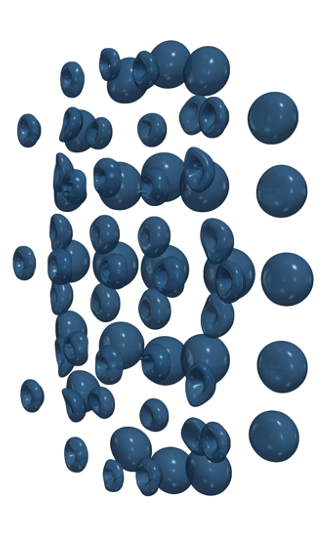}}; 
        \node (image) at (2in,0) {\includegraphics[width=1in]{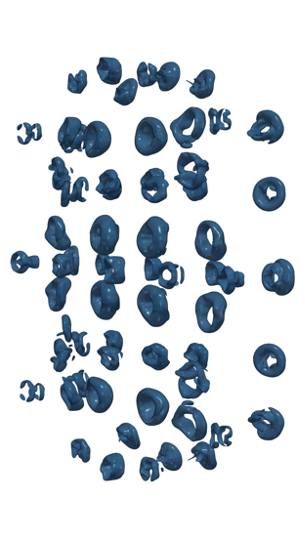}}; 
        \draw[ultra thick,-latex] (0,-1.05in) -- (2in, -1.05in);   
        \node[anchor=west] at (2in, -1.05in) {Time};
    \end{tikzpicture}
    \caption{
        A scale-resolved simulation of a collapsing bubble cloud.
        The bubble surface is shown at increasing points in time from left to right.
    }
    \label{fig:sb}
\end{figure}

Lastly, we show a shock bubble cloud interaction in water.
More specifically, the simulation shows a Mach~2.4 shock in water interacting with a cloud of 75~air~bubbles.
The simulation resolves 100~grid~cells in the bubble diameter at the initial condition and comprises 2~billion grid~cells.
It was advanced through 15.6~thousand time steps using 1024~MI250X GCDs in approximately 30~minutes. 
\Cref{fig:sb} shows the deformation of the bubbles at increasing times from left to right.

\section{Conclusion}
Directive-based hardware offloading is a simple way to develop portable codes for various GPU hardware while maintaining the ability to execute CPU code without modification.
We developed implementation and optimization strategies for efficiently using AMD and NVIDIA GPUs with OpenACC offloading.
Directive optimizations and manual inlining result in speedups of six and ten times.
Hardware tuned array packing with \texttt{hipBLAS} and compile time sizing of arrays in \texttt{private} directives on MI250X hardware yield an additional seven and thirty times speedup for select kernels.
Domain decomposition using 3D blocks results in weak scaling performance to 50\% of OLCF Summit with 97\% efficiency and 87\% of OLCF Frontier with 95\% efficiency.
Strong scaling efficiencies of 84\% and 81\% are observed when the device count is increased by 8 and 16 on NVIDIA V100 and AMD MI250X hardware.
The strong scaling efficiency of MI250X is increased to 91\% when GPU-aware MPI is used.

\section*{Acknowledgments}

SHB acknowledges support from US Office of Naval Research under grant numbers N00014-22-1-2519 (PM~Julie~Young) and N00014-24-1-2094 (PM~Chad~Stoltz) and the Department of Energy under DOE~DE-NA0003525 (Sandia~National~Labs) subcontract (PM~Ryan~McMullen), and hardware gifts from NVIDIA and AMD.
This research used resources of the Oak Ridge Leadership Computing Facility at the Oak Ridge National Laboratory, which is supported by the Office of Science of the U.S. Department of Energy under Contract No.~DE-AC05-00OR22725 (allocation CFD154, PI Bryngelson).
This work used Delta at the National Center for Supercomputing Applications through allocation PHY210084 (PI Bryngelson) from the Advanced Cyberinfrastructure Coordination Ecosystem: Services \& Support (ACCESS) program, which is supported by National Science Foundation grants \#2138259, \#2138286, \#2138307, \#2137603, and \#2138296.

\section*{Code Availability}
MFC is an open-source project available under the MIT license. 
Its source code is available at \href{https://github.com/MFlowCode/MFC}{github.com/MFlowCode/MFC}.
Additional information, documentation, and example simulations are available at \href{https://mflowcode.github.io}{mflowcode.github.io}.

\bibliographystyle{IEEEtranN}
\bibliography{main.bib}

\begin{thebibliography}{27}
\providecommand{\natexlab}[1]{#1}
\providecommand{\url}[1]{#1}
\csname url@samestyle\endcsname
\providecommand{\newblock}{\relax}
\providecommand{\bibinfo}[2]{#2}
\providecommand{\BIBentrySTDinterwordspacing}{\spaceskip=0pt\relax}
\providecommand{\BIBentryALTinterwordstretchfactor}{4}
\providecommand{\BIBentryALTinterwordspacing}{\spaceskip=\fontdimen2\font plus
\BIBentryALTinterwordstretchfactor\fontdimen3\font minus
  \fontdimen4\font\relax}
\providecommand{\BIBforeignlanguage}[2]{{%
\expandafter\ifx\csname l@#1\endcsname\relax
\typeout{** WARNING: IEEEtranN.bst: No hyphenation pattern has been}%
\typeout{** loaded for the language `#1'. Using the pattern for}%
\typeout{** the default language instead.}%
\else
\language=\csname l@#1\endcsname
\fi
#2}}
\providecommand{\BIBdecl}{\relax}
\BIBdecl

\bibitem[{De Vanna} and Baldan(2024)]{uranos20}
F.~{De Vanna} and G.~Baldan, ``{URANOS}-2.0: {I}mproved performance, enhanced
  portability, and model extension towards exascale computing of high-speed
  engineering flows,'' \emph{Computer Physics Communications}, vol. 303, p.
  109285, 2024.

\bibitem[Sathyanarayana et~al.(2023)Sathyanarayana, Bernardini, Modesti,
  Pirozzoli, and Salvadore]{streams2}
\BIBentryALTinterwordspacing
S.~Sathyanarayana, M.~Bernardini, D.~Modesti, S.~Pirozzoli, and F.~Salvadore,
  ``High-speed turbulent flows towards the exascale: {STREAmS}-2 porting and
  performance,'' 2023, {P}reprint. [Online]. Available:
  \url{https://arxiv.org/abs/2304.05494}
\BIBentrySTDinterwordspacing

\bibitem[Salvadore et~al.(2024)Salvadore, Rossi, Sathyanarayana, and
  Bernardini]{streamsOMP}
F.~Salvadore, G.~Rossi, S.~Sathyanarayana, and M.~Bernardini, ``Openmp offload
  toward the exascale using intel® gpu max 1550: evaluation of streams
  compressible solver,'' \emph{The Journal of Supercomputing}, vol.~80, no.~14,
  p. 21094–21127, Sep. 2024.

\bibitem[Bryngelson et~al.(2021)Bryngelson, Schmidmayer, Coralic, Meng, Maeda,
  and Colonius]{bryngelson2021}
S.~H. Bryngelson, K.~Schmidmayer, V.~Coralic, J.~C. Meng, K.~Maeda, and
  T.~Colonius, ``{MFC}: An open-source high-order multi-component, multi-phase,
  and multi-scale compressible flow solver,'' \emph{Computer Physics
  Communications}, vol. 266, p. 107396, 2021.

\bibitem[Wienke et~al.(2012{\natexlab{a}})Wienke, Springer, Terboven, and
  an~Mey]{OpenACC}
S.~Wienke, P.~Springer, C.~Terboven, and D.~an~Mey, ``Openacc --- {F}irst
  experiences with real-world applications,'' in \emph{Euro-Par 2012 Parallel
  Processing}, C.~Kaklamanis, T.~Papatheodorou, and P.~G. Spirakis, Eds.\hskip
  1em plus 0.5em minus 0.4em\relax Berlin, Heidelberg: Springer Berlin
  Heidelberg, 2012, pp. 859--870.

\bibitem[Andrianov and Warnecke(2004)]{andrianov2004}
N.~Andrianov and G.~Warnecke, ``The {R}iemann problem for the
  {B}aer--{N}unziato two-phase flow model,'' \emph{Journal of Computational
  Physics}, vol. 195, no.~2, pp. 434--464, 2004.

\bibitem[Saurel and Pantano(2018)]{Saurel2018}
R.~Saurel and C.~Pantano, ``\BIBforeignlanguage{en}{Diffuse-interface capturing
  methods for compressible two-phase flows},''
  \emph{\BIBforeignlanguage{en}{Annual Review of Fluid Mechanics}}, vol.~50,
  no.~1, p. 105–130, 2018.

\bibitem[Allaire et~al.(2002)Allaire, Clerc, and Kokh]{allaire2002}
G.~Allaire, S.~Clerc, and S.~Kokh, ``A five-equation model for the simulation
  of interfaces between compressible fluids,'' \emph{Journal of Computational
  Physics}, vol. 181, no.~2, pp. 577--616, 2002.

\bibitem[Le~Métayer and Saurel(2016)]{Metayer2016}
O.~Le~Métayer and R.~Saurel, ``The {N}oble--{A}bel stiffened-gas equation of
  state,'' \emph{Physics of Fluids}, vol.~28, no.~4, p. 046102, 2016.

\bibitem[Coralic and Colonius(2014)]{Coralic2014}
V.~Coralic and T.~Colonius, ``Finite-volume {WENO} scheme for viscous
  compressible multicomponent flows,'' \emph{Journal of Computational Physics},
  vol. 274, pp. 95--121, 2014.

\bibitem[Elwasif et~al.(2023)Elwasif, Godoy, Hagerty, Harris, Hernandez, Joo,
  Kent, Lebrun-Grandie, Maccarthy, Melesse~Vergara,
  et~al.]{elwasif2023application}
W.~Elwasif, W.~Godoy, N.~Hagerty, J.~A. Harris, O.~Hernandez, B.~Joo, P.~Kent,
  D.~Lebrun-Grandie, E.~Maccarthy, V.~Melesse~Vergara \emph{et~al.},
  ``Application experiences on a {GPU}-accelerated {Arm}-based {HPC} testbed,''
  in \emph{Proceedings of the HPC Asia 2023 Workshops}, 2023, pp. 35--49.

\bibitem[Radhakrishnan et~al.(2022)Radhakrishnan, {Le Berre}, and
  Bryngelson]{radhakrishnan22}
A.~Radhakrishnan, H.~{Le Berre}, and S.~H. Bryngelson, ``Scalable {GPU}
  accelerated simulation of multiphase compressible flow,'' in \emph{The
  International Conference for High Performance Computing, Networking, Storage,
  and Analysis (SC)}, Dallas, TX, USA, 2022, pp. 1--3.

\bibitem[Radhakrishnan et~al.(2024)Radhakrishnan, {Le Berre}, Wilfong, Spratt,
  {Rodriguez Jr.}, Colonius, and Bryngelson]{radhakrishnan24}
A.~Radhakrishnan, H.~{Le Berre}, B.~Wilfong, J.-S. Spratt, M.~{Rodriguez Jr.},
  T.~Colonius, and S.~H. Bryngelson, ``Method for portable, scalable, and
  performant {GPU}-accelerated simulation of multiphase compressible flow,''
  \emph{Computer Physics Communications}, vol. 302, p. 109238, 2024.

\bibitem[Toro(2009)]{toro2009riemann}
E.~Toro, \emph{Riemann {S}olvers and {N}umerical {M}ethods for {F}luid
  {D}ynamics: {A} {P}ractical {I}ntroduction}.\hskip 1em plus 0.5em minus
  0.4em\relax Springer Berlin Heidelberg, 2009.

\bibitem[Shu(2006)]{Shu2006}
C.-W. Shu, ``Numerical methods for hyperbolic conservation laws ({AM257}),''
  2006, lecture notes.

\bibitem[Vinokur(1983)]{VINOKUR1983}
M.~Vinokur, ``On one-dimensional stretching functions for finite-difference
  calculations,'' \emph{Journal of Computational Physics}, vol.~50, no.~2, pp.
  215--234, 1983.

\bibitem[Frigo and Johnson(2005)]{FFTW}
M.~Frigo and S.~G. Johnson, ``The design and implementation of {FFTW3},''
  \emph{Proceedings of the IEEE}, vol.~93, no.~2, pp. 216--231, 2005.

\bibitem[{NVIDIA Corporation}(2007)]{cuFFT}
\BIBentryALTinterwordspacing
{NVIDIA Corporation}, ``{cuFFT}: {H}igh-performance {CUDA} {FFT} library,''
  2007. [Online]. Available: \url{https://developer.nvidia.com/cufft}
\BIBentrySTDinterwordspacing

\bibitem[{A}dvanced Micro Devices~Inc.(2020)]{hipFFT}
\BIBentryALTinterwordspacing
{A}dvanced Micro Devices~Inc., ``{hipFFT}: {H}igh-performance fast {F}ourier
  transform library,'' 2020, {Github} Repository. [Online]. Available:
  \url{https://github.com/ROCmSoftwarePlatform/hipFFT}
\BIBentrySTDinterwordspacing

\bibitem[Miller(2022)]{millersilo}
\BIBentryALTinterwordspacing
M.~Miller, ``Silo--{A} mesh and field {I/O} library and scientific database,''
  \emph{Lawrence Livermore National Laboratory}, 2022, {Github} repository.
  [Online]. Available: \url{https://github.com/LLNL/Silo}
\BIBentrySTDinterwordspacing

\bibitem[Ahrens et~al.(2005)Ahrens, Geveci, and Law]{Ahrens2005ParaViewAE}
J.~P. Ahrens, B.~Geveci, and C.~C. Law, ``Paraview: An end-user tool for
  large-data visualization,'' in \emph{The Visualization Handbook}, 2005.

\bibitem[Childs et~al.(2012)Childs, Brugger, Whitlock, Meredith, Ahern,
  Pugmire, Biagas, Miller, Harrison, Weber, Krishnan, Fogal, Sanderson, Garth,
  Bethel, Camp, R\"{u}bel, Durant, Favre, and Navr\'{a}til]{VisIt}
H.~Childs, E.~Brugger, B.~Whitlock, J.~Meredith, S.~Ahern, D.~Pugmire,
  K.~Biagas, M.~Miller, C.~Harrison, G.~H. Weber, H.~Krishnan, T.~Fogal,
  A.~Sanderson, C.~Garth, E.~W. Bethel, D.~Camp, O.~R\"{u}bel, M.~Durant, J.~M.
  Favre, and P.~Navr\'{a}til, ``Visit: An end-user tool for visualizing and
  analyzing very large data,'' in \emph{High Performance
  Visualization--Enabling Extreme-Scale Scientific Insight}, October 2012, pp.
  357--372.

\bibitem[Wienke et~al.(2012{\natexlab{b}})Wienke, Springer, Terboven, and
  an~Mey]{wienke2012}
S.~Wienke, P.~Springer, C.~Terboven, and D.~an~Mey, ``{OpenACC} --- {F}irst
  experiences with real-world applications,'' in \emph{Euro-Par 2012 Parallel
  Processing}, C.~Kaklamanis, T.~Papatheodorou, and P.~G. Spirakis, Eds.\hskip
  1em plus 0.5em minus 0.4em\relax Berlin, Heidelberg: Springer Berlin
  Heidelberg, 2012, pp. 859--870.

\bibitem[{NVIDIA Corporation}(2019)]{cuTENSOR}
\BIBentryALTinterwordspacing
{NVIDIA Corporation}, ``{cuTENSOR}: {A} high-performance {CUDA} library for
  tensor primitives,'' 2019. [Online]. Available:
  \url{https://developer.nvidia.com/cutensor}
\BIBentrySTDinterwordspacing

\bibitem[{A}dvanced {M}icro~{D}evices {I}nc.(2020)]{hipBLAS}
\BIBentryALTinterwordspacing
{A}dvanced {M}icro~{D}evices {I}nc., ``{hipBLAS}: {H}igh-performance basic
  linear algebra subprograms library,'' 2020, {Github} Repository. [Online].
  Available: \url{https://github.com/ROCmSoftwarePlatform/hipBLAS}
\BIBentrySTDinterwordspacing

\bibitem[Chandrasekaran and Juckeland(2017)]{sunita2017}
S.~Chandrasekaran and G.~Juckeland, \emph{{OpenACC} for {P}rogrammers:
  {C}oncepts and {S}trategies}, 1st~ed.\hskip 1em plus 0.5em minus 0.4em\relax
  Addison-Wesley Professional, 2017.

\bibitem[Aradi(2021)]{fypp}
\BIBentryALTinterwordspacing
B.~Aradi, ``{Fypp}: {P}ython-powered {F}ortran metaprogramming,'' 2021,
  {GitHub} Repository. [Online]. Available: \url{https://github.com/aradi/fypp}
\BIBentrySTDinterwordspacing

\end{thebibliography}
\vspace{12pt}

\end{document}